# Combustion Phasing Modelling and Control for Compression Ignition Engines with High Dilution and Boost Levels

Wenbo Sui[1], Carrie M. Hall[1]*

1:  Illinois Institute of Technology, Department of Mechanical, Materials, and Aerospace Engineering

*Email: chall9@iit.edu

## Abstract:

Because fuel efficiency is significantly impacted by the timing of combustion in internal combustion engines, accurate control of combustion phasing is critical. In this paper, a nonlinear combustion phasing model is introduced and calibrated, and both a feedforward model-based control strategy and an adaptive model-based control strategy are investigated for combustion phasing control. The combustion phasing model combines a knock integral model, burn duration model and a Wiebe function to predict the combustion phasing of a diesel engine. This model is simplified to be more suitable for combustion phasing control and is calibrated and validated using simulations and experimental data that include conditions with high exhaust gas recirculation fractions and high boost levels. Based on this model, an adaptive nonlinear model-based controller is designed for closed-loop control, and a feedforward model-based controller is designed for open-loop control. These two control approaches were tested in simulations. The simulation results show that during transient changes the CA50 (the crank angle at which 50% of the mass of fuel has burned) can reach steady state in no more than 5 cycles and the steady state errors are less than ±0.1 crank angle degree (CAD) for adaptive control, and less than ±0.5 CAD for feedforward model-based control.

## I.  Introduction

Fuel efficiency demands and emissions requirements have continued to increase for diesel engines, which are the dominant power source for mid-duty and heavy-duty automotive vehicles. Optimal combustion phasing is essential for maintaining high efficiencies and as such, control of this parameter is of much interest [1-4]. As shown in Fig. 1, modern diesel engines typically are equipped with a number of different technologies including turbochargers and exhaust gas recirculation systems. On such engines, the fresh air is often compressed by a compressor and mixed with recirculated exhaust gas in the intake manifold. After the intake process, diesel is injected into the cylinders, and combustion occurs. The exhaust gas exits the cylinders to the exhaust manifold after combustion. A portion of the exhaust flow can be routed through an exhaust gas recirculation (EGR) loop, while the remainder can enter the turbine and drive the compressor.

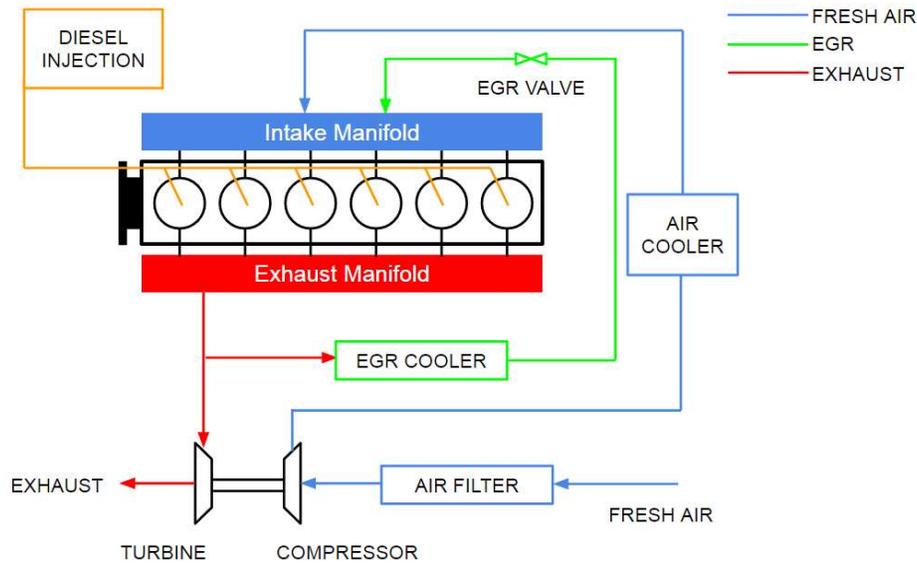

Figure 1. Schematic of diesel engine system

With the inclusion of technologies like EGR, variable geometry turbochargers (VGT), diesel engines are becoming increasingly complicated. As such, simple rule-based control strategies or methods based on look-up tables [5, 6] are not always satisfactory on these more complex engines and the automotive industry has turned to model-based and closed loop control techniques on modern engines [7, 8]. These model-based control strategies have been used to accurately control engine processes including air flows, in-cylinder gas composition and combustion timing. To apply these strategies for combustion control, precise combustion phasing models are required and therefore, have been a topic of much focus.

A variety of combustion modeling and estimation techniques have been proposed over the years. In [9-12], Arrhenius type models were used to estimate the start of combustion (SOC) with a reasonable accuracy. However, such models typically require in-cylinder oxygen and fuel concentration measurements and these are not available on production engines. As such, strategies that leverage Arrhenius type models would require additional sensors or additional estimators. The Shell auto-ignition method has also been employed in [13-15] to predict SOC with good precision. However, a detailed CFD (computational fluid dynamics) model is needed for the calibration of this model since many parameters in this model are calibrated based on chemical kinetics. Thus, using a Shell auto-ignition model can be challenging in real applications.

Another way to capture part of the combustion phasing is with a knock integral model (KIM), which was first proposed by Livengood and Wu. This model was first used to predict the knock of a spark ignition engine [16], but can also be utilized to predict SOC on diesel engines [17-19]. In [17], Hillion et al. employed a KIM to predict SOC and the end of the cool flame on a diesel engine. The authors of [18] and [19] have used a KIMs not only to predict the SOC, but also to estimate the emissions including CO and NO mass flow rate. In addition, the KIM has been modified in work led by Shahbakhti to include the impact of varying air-fuel ratio and EGR fraction on SOC in a homogenous charge compression ignition (HCCI) engine [20]. Different forms of the Arrhenius factor that is included in the knock integral were also studied in [21] in order to predict and control the SOC and location of peak pressure in a HCCI engine. More recently, combustion phasing estimation techniques have also been developed that utilize engine speed measurements along with some estimate or measurement of torque to predict CA50 [22, 23].

In addition to modeling diesel engine combustion phasing, there are also many studies that focus on the control of combustion phasing for diesel engines including [17], [24], [25], [26], [27] and [28]. Model-based control of combustion phasing was studied in [17]; however, the CA50 reached steady state after more than 20 cycles when the operation condition changed in this study. Willems et al. and Yu et al. employed PI and PID feedback controllers to control the CA50 and IMEP with in-cylinder pressure sensors [24] and [25]. Meanwhile, PI feedback control based on the CA50 estimated by crankshaft torque measurements was used in [26] and a fuzzy logic controller was used with an ion current signal to control the location of peak premixed combustion (LPPC) was studied in [27]. While these control approaches worked well, most commercial diesel engines do not have the in-cylinder pressure sensors, crankshaft torque sensors or ion current sensors required for these feedback control techniques. As such, these feedback strategies are still hard to apply in production engines. In [28], an adaptive feedforward control based on a RBFNN (radial basis function neural network) is proposed to control the start of combustion and has a much lower settling time than with PI control. Feedforward strategies have also been explored in [29], but the underlying model assumed complete combustion and required an iterative procedure to find the needed injection timing. A feedforward controller with an adaptive parameter estimator also used by Larimore et. al. [30] to control CA50 on HCCI engines. Larimore used an online adaptive estimator to adapt parameters in an engine combustion model and accurately predict CA50 on engines using an HCCI strategy.

In contrast to these prior studies, this work focuses on modeling SOC as well as CA50 for a high efficiency diesel engine and introduces two control approaches that leverage this model. Many prior studies have focused on controlling SOC, but since combustion efficiency is tied to an optimal CA50 timing, control of CA50 more expedient [4, 31]. While controlling SOC or other combustion metrics like LPPC is helpful, it may not provide optimal fuel efficiency for diesel engines. This work predicts both SOC and CA50 in order to allow for control of combustion phasing. The model predicts SOC using a modified knock integral model (MKIM), but also includes a separate burn duration (BD) model, and calculates CA50 based on the predicted SOC and BD. The integral model is simplified into a nonlinear model that is calibrated by more than 500 simulations and validated against simulations and experimental data. In this work, only a single fuel pulse is considered. Pilot injections or a more novel rate shape could impact the accuracy of the developed model but these strategies were outside the scope of this work.

The main novelties of this work are 1) the application of accepted combustion modeling techniques to conditions encountered in high efficiency diesels where combustion is mixing controlled and high EGR rates and boost pressures are used, 2) a simplification method that allows the integral term in the knock integral model to be simplified and makes model-based control more feasible, and 3) consideration of the suitability of such a simplified model for diesel combustion control. In the knock integral model, which is used to predict SOC, the integral term contains dynamic in-cylinder pressure and temperature, which is hard to be measure for production engines. Prior efforts including those in [20] and [26] have used the pressure and temperature at IVC to calculate the dynamic pressure and temperature, but this requires an iterative solution for SOI. In this paper, the pressure and temperature at SOI are used instead of those dynamic pressure and temperature to simplify the knock integral model. This simplified version also gives accurate results and its utility for control applications is explored in this work in two different control methods.

An adaptive feedback controller is designed to control the CA50 based on the nonlinear model and relies on a measurement of CA50. Although there are some commercial diesel engines that have in-cylinder sensors that could provide CA50 feedback, most production engines do not. As such, this adaptive feedback control strategy may not be useful for many production engines, but illustrates a "best case" scenario. To overcome this disadvantage of the feedback controller, another control strategy that uses feedforward model-based control is also studied.

The paper details the development of the combustion phasing prediction model and its simplification. After the model development is discussed, model calibration and validation using simulation and experimental data are given. The two different control methodologies are then introduced and the simulation results with these control strategies are presented and discussed. Finally, some conclusions are made regarding the performance of the control techniques used in this work.

## II. Modeling Diesel Engine Combustion Phasing

In order to ultimately control combustion phasing via model-based control methods, a control-oriented model of combustion phasing is required. In this work, a modified knock integral model (MKIM) is derived to predict the SOC. Afterward, a burn duration model that accounts for dilution fraction is developed. Then, CA50 is modeled using a Wiebe function. Finally, the submodels are combined and simplified to a nonlinear model. The details of each model are discussed in this section.

### A. SOC Model

Knock integral models (KIMs) are widely used to predict the SOC for different engines [16-20]. In KIMs, the relationship between SOI, SOC and an Arrhenius function is expressed as

$$\int_{SOI}^{SOC} \frac{\tau}{N} d\theta = 1 \tag{1}$$

where *SOI* is the crank angle of start of fuel injection, *SOC* represents the crank angle of start of combustion, $N$ denotes the engine speed, and $\tau$ is the Arrhenius function. The Arrhenius function can take different forms but generally is a function of compression temperatures and pressures. Here the Arrhenius function is given by

$$\tau = \frac{1}{c_1 EGR + c_2} \phi^{c_3} \exp\left(-\frac{c_4 P^{c_5}}{T}\right) \tag{2}$$

where $EGR$ represents the EGR fraction, $\phi$ is the fuel equivalence ratio, $T$ is the temperature, $P$ is pressure, and $c_1, c_2, c_3, c_4$ and $c_5$ are constants.
The equivalence ratio is defined as

$$\phi = \left(\frac{m_{\text{fuel}}}{m_{\text{air}}}\right) / \left(\frac{m_{\text{fuel}}}{m_{\text{air}}}\right)_{st} \tag{3}$$

where $m_{\text{fuel}}$ is the diesel injection mass and $m_{\text{air}}$ is the mass of air in the engine cylinder. The subscript $st$ indicates stoichiometric conditions.

Substituting Eqn. (2) into Eqn. (1), the integral equation can be rewritten as

$$\int_{SOI}^{SOC} \frac{\phi^{c_3} \exp\left(-\frac{c_4 P^{c_5}}{T}\right)}{(c_1 EGR + c_2) N} d\theta = 1. \tag{4}$$

In this knock integral model, an assumption has been made that combustion does not occur during the compression and auto-ignition phases. As such, heat losses from SOI to SOC can be ignored [32].

## B. Burn Duration Model

Proper timing of CA50 is pivotal to maximizing efficiency. In order to predict CA50, this work leverages a burn duration model. Burn duration is defined as the period from CA10 (the crank angle at which 10% of the mass of fuel has burned) to CA90 (the crank angle at which 90% of the mass of fuel has burned). Burn duration is described by

$$BD = c_6 (1 + X_d)^{c_7} \phi^{c_8} \tag{5}$$

where $X_d$ is the dilution fraction, $\phi$ indicates the equivalence ratio, and $c_6, c_7$ and $c_8$ are constant parameters. The dilution fraction can be given by

$$X_d = EGR + X_r \tag{6}$$

where $EGR$ is the EGR fraction, and $X_r$ represents the residual fraction. The residual fraction $X_r$ is defined in Eqn. (7).

$$X_r = \frac{m_r}{m_{air} + m_{fuel} + m_{egr}} \tag{7}$$

In Eqn. (7), $m_r$ is the mass of residual gas in the cylinder, $m_{air}$ is the mass of air entering the cylinder, $m_{fuel}$ is mass of diesel injected into the cylinder and $m_{egr}$ is the mass of EGR gas entering the cylinder. Based on knowledge of $EGR$, $X_r$ and $\phi$ (from measurements or underlying estimators), the BD model can be calibrated and used for predictions.

## C. CA50 Prediction Model

With the prediction of SOC and BD, CA50 can be estimated by using a Wiebe function [1,2]. The Wiebe function has the form

$$x_b(\theta) = 1 - \exp\left(-a\left[\frac{\theta - SOC}{BD}\right]^b\right) \tag{8}$$

where $x_b$ is the mass fraction of burned fuel, $SOC$ is the crank angle of SOC, $BD$ indicates the burn duration, and $a$ and $b$ are constant coefficients.

CA50 is found by considering the point when $x_b$ in Eq. (8) equals 0.5. Thus, Eqn. (8) can be rewritten as

$$0.5 = 1 - \exp\left(-a\left[\frac{CA50 - SOC}{BD}\right]^b\right) \tag{9}$$

and rearranged to yield

$$\left[\frac{CA50 - SOC}{BD}\right]^b = \frac{\ln 2}{a} \tag{10}$$

Taking the $b$th root of both sides, the equation can be put in the form

$$\frac{CA50 - SOC}{BD} = \left[\frac{\ln 2}{a}\right]^{1/b} \tag{11}$$

and simplified to describe CA50 as

$$CA50 = SOC + \left[\frac{\ln 2}{a}\right]^{1/b} BD. \tag{12}$$

Substituting Eqn. (5) into Eqn. (12), CA50 can be found by:

$$CA50 = SOC + c_9(1 + X_d)^{c_7} \phi^{c_8} \tag{13}$$

in which $c_9$ is given by:

$$c_9 = \left[\frac{\ln 2}{a}\right]^{1/b} c_6. \tag{14}$$

With these models, SOC can be predicted by the MKIM in Eqn. (4), and CA50 can be estimated by Eqn. (14).

## D. CA50 Model Simplification

Although Eqns. (4) and (14) can be used to estimate SOC and CA50, the integral term in Eqn. (4) makes it difficult to apply this combustion dynamic model in combustion phasing control efforts. Therefore, a simplification is needed before considering control design. In order to simply this model, equivalence ratio, engine speed and EGR fraction will be treated as constant during a cycle. As such, these terms can be taken out from the integral term transforming Eq. (4) to

$$\frac{\phi^{c_3}}{(c_1 EGR + c_2)N} \int_{SOI}^{SOC} \exp\left(-\frac{c_4 P^{c_5}}{T}\right) d\theta = 1. \tag{15}$$

The pressure and temperature in Eqn. (15) can be related to the conditions at intake valve closing (IVC) using a polytropic relationship and expressed as

$$T = T_{IVC} \left(\frac{V_{IVC}}{V(\theta)}\right)^{k_c - 1} \tag{16}$$

$$P = P_{IVC} \left(\frac{V_{IVC}}{V(\theta)}\right)^{k_c} \tag{17}$$

where $T_{IVC}$, $P_{IVC}$ and $V_{IVC}$ are the temperature, pressure and cylinder volume at IVC, $k_c$ represents the polytropic constant and $V(\theta)$ is the cylinder volume at crank angle $\theta$. Typically, the SOI occurs from -10° aTDC to 5° aTDC, and the SOC occurs only 1 – 5 CAD after SOI. As such, the change in cylinder volume during this time period is small. Based on Eqns. (16) and (17), the temperature and pressure do not change drastically from SOI to SOC. To simplify the integral, the temperature at SOI and pressure at SOI can be used instead of the dynamic temperature and pressure in the exponential term in Eqn. (15). Because the temperature and pressure at SOI are constant values during a specific cycle, the exponential term can be treated as constant during that period. Thus, the integral term can be simplified to

$$\frac{\phi^{c_3}}{(c_1 EGR + c_2)N} \exp\left(-\frac{c_4 P_{SOI}^{c_5}}{T_{SOI}}\right)(SOC - SOI) = 1. \tag{18}$$

Eqn. (18) can be rearranged to yield

$$SOC = SOI + (c_1 EGR + c_2)N\phi^{-c_3} \exp\left(\frac{c_4 P_{SOI}^{c_5}}{T_{SOI}}\right). \tag{19}$$

Substituting Eqn. (19) into Eqn. (13), CA50 is given by:

$$CA50 = SOI + (c_1 EGR + c_2)N\phi^{-c_3} \exp\left(\frac{c_4 P_{SOI}^{c_5}}{T_{SOI}}\right) + c_9(1 + X_d)^{c_7} \phi^{c_8}. \tag{20}$$

Thus, CA50 can be estimated by Eqn. (20) with EGR fraction, engine speed, equivalence ratio, average pressure, average temperature and dilution fraction.

## III. CA50 Model Validation

The CA50 model has several constants that must be calibrated. This was accomplished using data from an engine test cell along with a higher fidelity simulation software as discussed in this section.

## A. Diesel Engine Setup

Data from a 2010 Navistar Maxxforce 13 heavy-duty engine was used in the calibration efforts. This engine is a six-cylinder 12.4L diesel engine, equipped with a variable geometry turbocharger (VGT) and cooled EGR. The engine specifications are shown in Table 1.

Table 1. Engine Specifications

| | |
|---|---|
| Displacement Volume | 12.4L |
| Number of Cylinders | 6 |
| Compression Ratio | 17:1 |
| Valves per cylinder | 4 |
| Bore | 126mm |
| Stroke | 166mm |
| Connecting Rod length | 251mm |
| Diesel Fuel System | 2200 bar common rail |
| Air System | 2-stage turbocharger |

Experiments were conducted at 10 different engine operating conditions and are discussed in more detail in [33]. The ranges of these operating conditions are shown in Table 2, and the speed-load map of these operating points is also shown in Fig. 2. These points do not cover the entire speed load range of the engine, but are in a region in which high EGR rates and boost pressure can be leveraged to produce high efficiencies. The wide range of EGRs considered here are not typical over the entire operating region and the combustion model developed is particularly suited for these conditions.

All operating conditions consider a single pulse fuel injection. The injection duration does not vary significantly across the different load conditions, but the injection pressures change dramatically from 900 bar to 11,000 bar as the load is increased. Some of the values are beyond the range common in industry, because the base experiments included high injection pressures in an effort to find highly efficiency operating conditions. Most of the injection pressures are from 1,500 bar to 6,000 bar in the simulations. With the single pulse injection, there are some conditions under which injection and combustion overlap.

Table 2. Range of Parameters in Experimental Data

| Quantity | Minimum Value | Maximum Value |
|---|---|---|
| Diesel Injection Quantity (mg) | 56.8 | 193.5 |
| Engine Speed (RPM) | 1200 | 1500 |
| Intake Manifold Temperature (K) | 309.5 | 325 |
| Intake Manifold Pressure (bar) | 1.87 | 4.07 |
| Diesel Equivalence Ratio (-) | 0.2971 | 0.5418 |
| Exhaust Gas Recirculation (%) | 3.9 | 37.8 |
| Brake Mean Effective Pressure (bar) | 5.5 | 17.4 |
| Intake Valve Closing (° aTDC) | -148.5 | -148.5 |
| Intake Valve Opening (° aTDC) | -363.5 | -363.5 |
| Exhaust Valve Opening (° aTDC) | 137 | 137 |
| Exhaust Valve Closing (° aTDC) | 389 | 389 |

aTDC: after top dead center

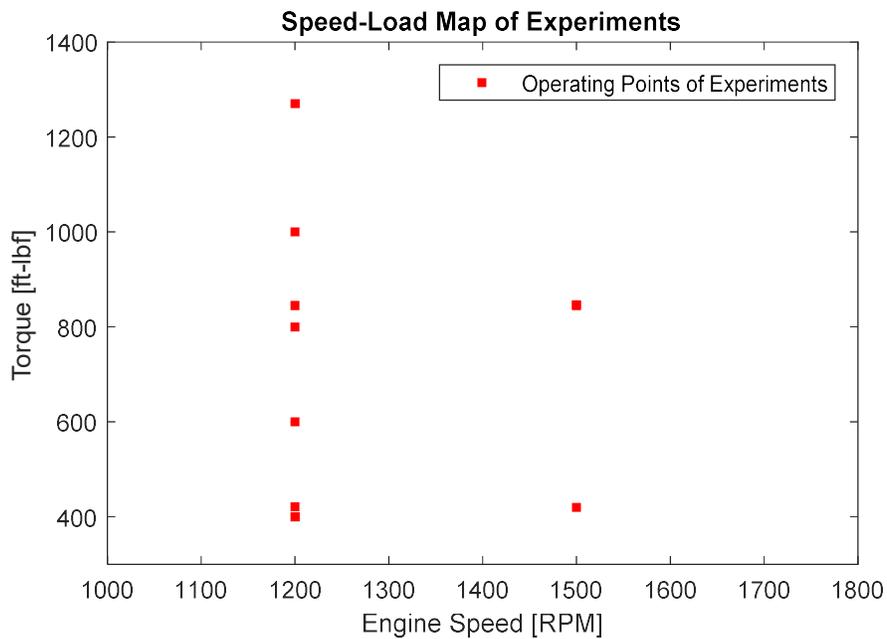

Figure 2. Speed-load map of experimental operating points

In order to consider additional cases and control methods, a simulation model was developed in Gamma Technology's Integrated Simulation Environment (GT-ISE) and validated against experimental data as discussed in [34]. This model simulates the combustion process based on a combination of the DI-Pulse and SI-Turb predictive models. The simulation model was calibrated with 10 experimental diesel-only cases as well as 10 dual-fuel cases. A comparison between the experimental and simulated CA50s for the diesel cases is given in Table 3 and demonstrates that the GT-power model can accurately predict the CA50 at different operating points. The GT-power simulation model is able to predict CA50 for the Navistar engine with an uncertainty of ±0.84 CAD.

Table 3. Comparison of Experimental and Simulated CA50

| Case Number | Engine Speed (rpm) | Air-Fuel Ratio (-) | Intake Manifold Pressure (bar) | EGR (%) | Experimental CA50 (dATDC) | Simulated CA50 (dATDC) |
|---|---|---|---|---|---|---|
| 1 | 1200 | 30.4 | 3.09 | 35.8 | 12.4 | 11.58 |
| 2 | 1200 | 41.1 | 1.87 | 37.8 | 12.4 | 11.86 |
| 3 | 1200 | 26.6 | 4.97 | 36.8 | 14.1 | 13.58 |
| 4 | 1500 | 29.8 | 2.29 | 37.4 | 13.5 | 12.43 |
| 5 | 1500 | 30.7 | 3.23 | 30.9 | 14.1 | 12.85 |
| 6 | 1500 | 27.9 | 3.12 | 36.7 | 14.8 | 13.03 |
| 7 | 1200 | 45.3 | 2.58 | 12.3 | 10.5 | 11.02 |
| 8 | 1200 | 38.6 | 2.56 | 6.89 | 11.2 | 11.19 |
| 9 | 1200 | 32.2 | 2.55 | 3.90 | 12.3 | 11.80 |
| 10 | 1200 | 48.7 | 2.57 | 17.2 | 9.9 | 10.98 |

This validated GT-ISE model was used to perform diesel engine combustion simulations over a range of operating conditions, as shown in Table 4. A total of 516 simulations were performed and the results were utilized to calibrate the parameters in the combustion phasing model.

Table 4. Range of Parameters in Simulations

| Quantity | Minimum Value | Maximum Value |
|---|---|---|
| Engine Speed (RPM) | 1200 | 1500 |
| $T_{IVC}$ (K) | 372.6 | 413.9 |
| $P_{IVC}$ (bar) | 2.85 | 4.38 |
| Diesel Equivalence Ratio (-) | 0.5 | 0.9 |
| Exhaust Gas Recirculation (%) | 0 | 50 |
| Start of Injection (° aTDC) | -5 | 5 |

aTDC: after top dead center

## B. Calibration of CA50 Prediction Model

The parameters in the CA50 model were calibrated based on the GT-ISE simulations. These simulations were run with a step size of 0.1 CAD. The root mean square error (RMSE) of CA50 was minimized by a batch gradient descent algorithm for the CA50 prediction model. All 516 simulations were utilized and the model calibration procedure is shown in Fig. 3. At the beginning of the calibrations, an initial guess was given for the parameters and the gradient of the RMSE is set to zero. For each simulation result, the CA50 was estimated based on those parameters, and the error was calculated. Next, the gradient of the error was computed, and it was added to the gradient of RMSE. The error is evaluated for all simulations and once all simulations were tested, the parameters were updated by the gradient of the RMSE. The calibration iteration stopped updating the parameters when the RMSE reached a steady value and could no longer decrease.

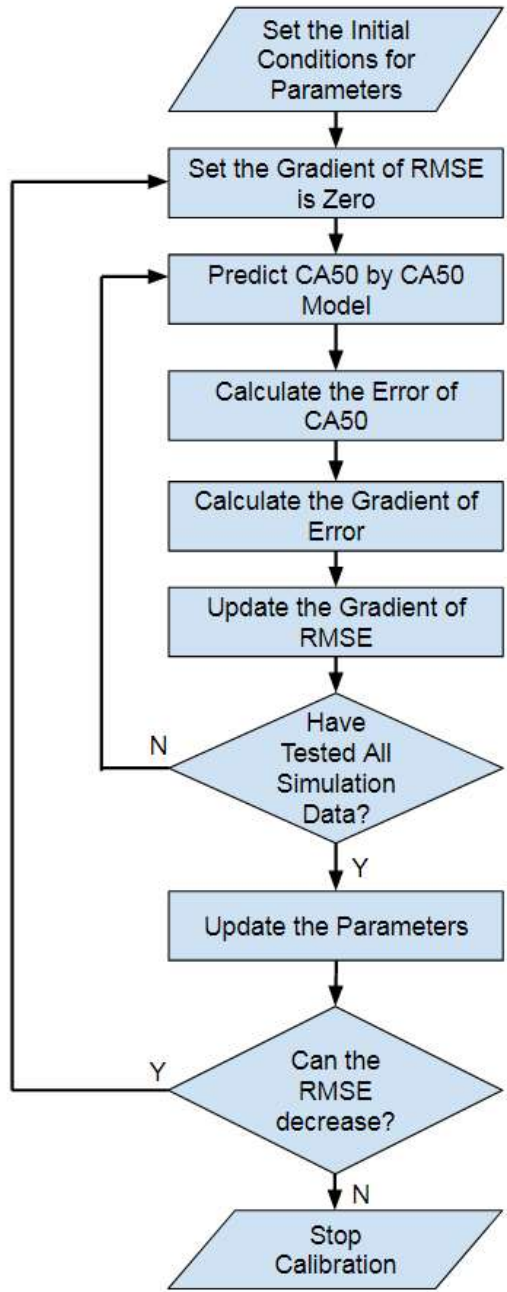

Figure 3. Model calibration procedure

The optimized parameters of the CA50 prediction model in Eqn. (20) are given in Table 5.

Table 5.    Parameters of CA50 Prediction Model

| | |
|---|---|
| $c_1$ | $2.000 \times 10^{-6}$ |
| $c_2$ | $2.705 \times 10^{-6}$ |
| $c_3$ | -0.128 |
| $c_4$ | 10643.118 |
| $c_5$ | -0.312 |
| $c_7$ | 0.371 |
| $c_8$ | 0.0165 |
| $c_9$ | 4.784 |
| $k_c$ | 1.176 |

## C. Validation of SOC and CA50 Prediction

With the parameters shown in Table 5, SOC can be estimated by Eqn. (19), and CA50 can be predicted by Eqn. (20). The SOC and CA50 prediction results are compared to all 516 data points as well as experimental results. The comparison with the GT simulation model is shown in Figs. 4 and 5. In Fig. 4, the x axis is the SOC from GT-ISE simulation, and the y axis is the SOC predicted from MKIM. The dashed blue lines show ±1 CAD error limits. The standard deviation of the prediction error is 0.2972 CAD, and the maximum error is 0.9695 CAD.

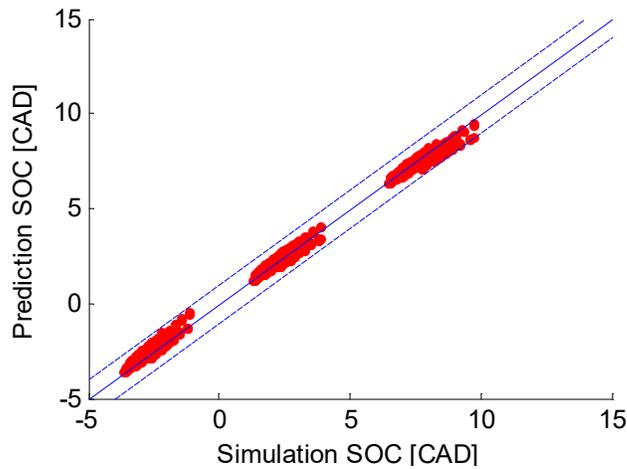

Figure 4. Comparison of SOC prediction and GT simulations

Some SOCs are over or under predicted and this is largely because the MKIM model ignores wall heat losses. As the SOI is delayed or the intake manifold temperature increases, the heat lost to the walls increases causing the MKIM to have larger errors. Error is also introduced since the temperature and pressure at SOI are utilized instead of the dynamic temperature and pressure. However, the errors are still relatively low for all operating conditions.

Fig. 5 demonstrates the accuracy of the CA50 prediction by comparing the CA50 from GT simulations on the x axis with the CA50 predicted from the model on the y axis. The dashed blue lines again represent ±1 CAD errors. The standard deviation of the CA50 prediction error is 0.3255 CAD, and the maximum error is 0.8711 CAD. Comparison with Fig. 4 shows that the main inaccuracies in the CA50 prediction are inherited from errors in the SOC prediction.

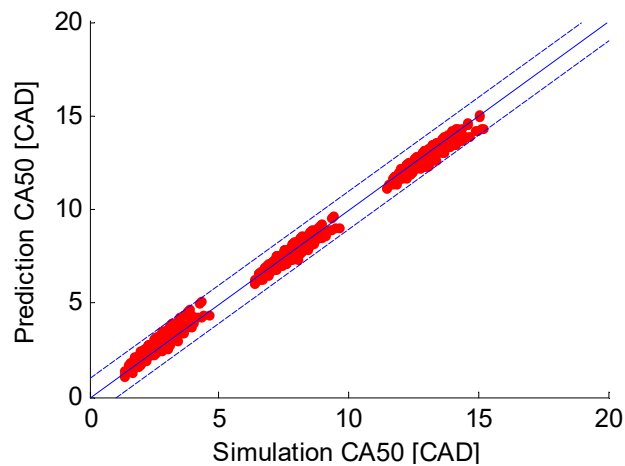

Figure 5. Comparison of CA50 prediction and GT simulations

The models for SOC and CA50 were also tested against the original experimental data. The SOCs and CA50s from the experimental data were calculated based on the in-cylinder pressure and volume trace using GT-power's three pressure analysis. The results are given in Figs. 6 and 7. The standard deviation of the SOC prediction error is 0.2060 CAD, and the maximum error is 0.4858 CAD. In Fig. 7, the standard deviation error of the CA50 predictions is 0.2199 CAD, and the maximum error is 0.7630 CAD. Together with the comparison to the more detailed simulation results, Figs. 6 and 7 show that the combustion phasing model can estimate SOC and CA50 with reasonable accuracy.

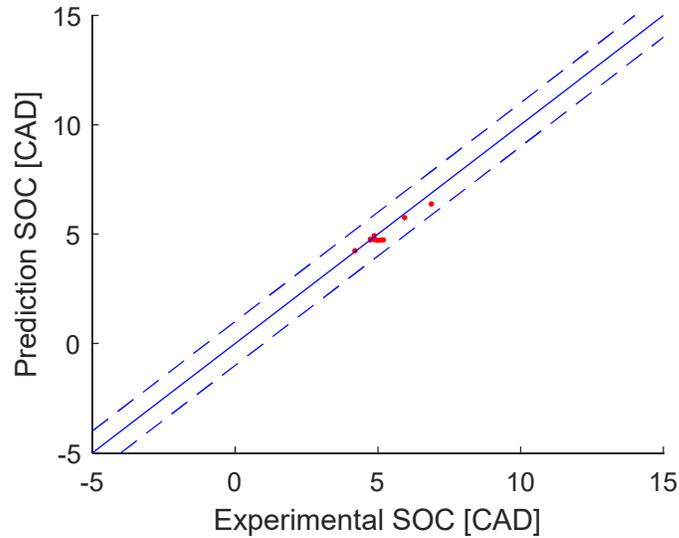

Figure 6. Comparison of SOC prediction and experimental data

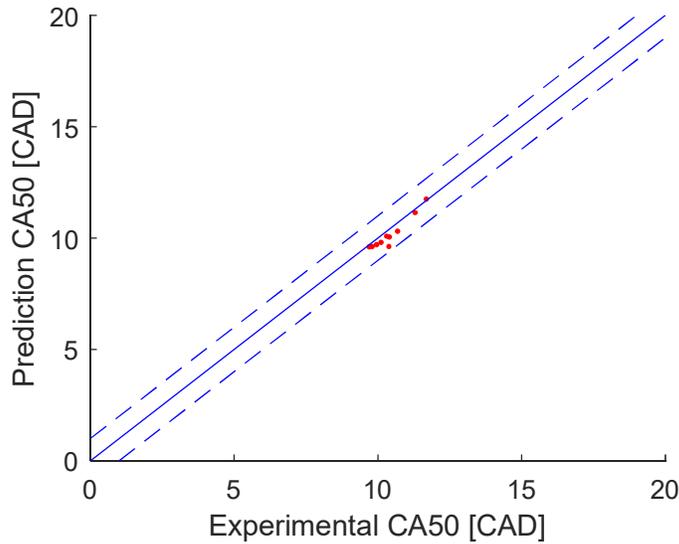

Figure 7. Comparison of CA50 prediction and experimental data

The results from Figs. 4 to 7 demonstrate that the models can estimate SOC and CA50 accurately at the steady state conditions, but the performance during transient conditions is also of concern. The prediction accuracy was also studied during transient conditions in which cycle-to-cycle variations may detrimentally affect the performance. Three transient simulations were tested. In these simulations, the start of fuel injection is 0 CAD, the average intake manifold pressure is 2 bar, and the average intake manifold temperature is 300 K.

In the first test, an engine speed transient was considered. EGR fraction was 25%, and the equivalence ratio was 0.7 during the 10 seconds simulation. The engine speed was 1200 RPM during the first 5 seconds, smoothly changed to 1500 RPM over 0.5 seconds, and stayed at 1500 RPM until the end of simulation. The CA50 prediction for this case is shown in Fig. 8a and the maximum error is -1.08 CAD. During the majority of the simulation, the absolute value of prediction error is less than 0.5 CAD.

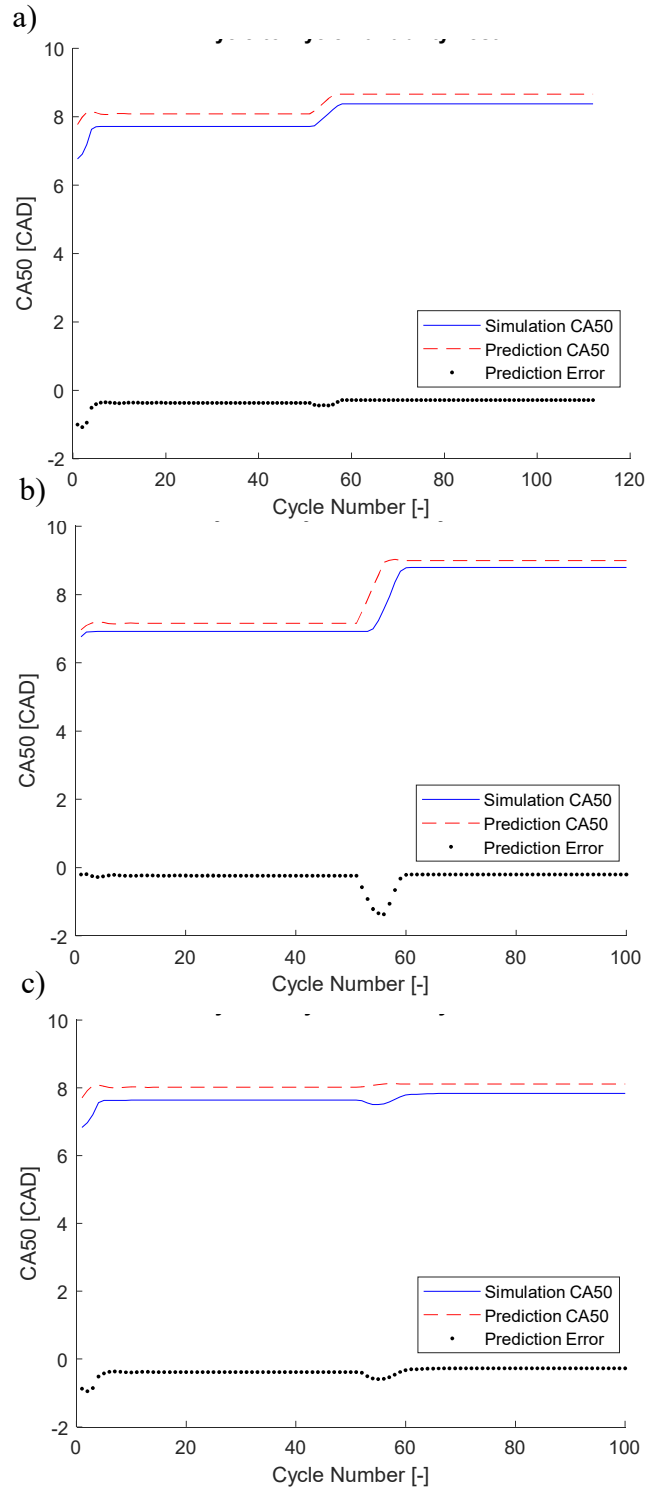

Figure 8. Prediction accuracy during an a) engine speed transient, b) EGR fraction transient, and c) equivalence ratio transient

Next, a change in EGR fraction was considered. Engine speed was held at 1200 RPM, and the equivalence ratio was 0.7 in this test. In the first 5 seconds, the EGR fraction was 0, but increases to 50% over 0.5 seconds at the 5 second mark. As shown in Figure 8b, the absolute value of transient prediction error is less than 1.5 CAD for this case even during a large change in EGR.

Lastly, the model prediction error during an equivalence ratio change was evaluated. During this transient, the engine speed was 1200 RPM, and the EGR fraction was 25%. The equivalence ratio was 0.5 before the transient starts at 5 seconds, and changed to 0.9 during a 0.5 second period. The results of this case are shown in Figure 8c. The prediction error is small even during the transient with a maximum error of 0.95 CAD. These results indicate that the CA50 model accurately predicts CA50 in steady state conditions as well as during transients.

In the next section, these models will be used in two control approaches, one which is adaptive and assumes a measurement of CA50 is available and one which uses a feedforward control approach and does not rely on a measured CA50.

## IV. Adaptive Controller Design

### A. State-Space Model Derivation

The dynamic model of CA50 expressed in Eqn. (20) can also be put in state space form for control purposes. In this work, the fuel injection timing or SOI is set as the control signal and CA50 is the target to track. Therefore, SOI is employed as the input of the system $u$, and CA50 is the output of the system $y$. The exponential term and the dilution fraction term are used as the states. As such, a state space version of the CA50 model can be given as:

$$y = u + \alpha x_1 + \beta x_2 \tag{21}$$

$$x_1 = (c_1 EGR + c_2) \exp\left(\frac{c_4 P_{SOI}^{c_5}}{T_{SOI}}\right) \tag{22}$$

$$x_2 = c_9(1 + X_d)^{c_7} \tag{23}$$

$$\alpha = N\phi^{-c_3} \tag{24}$$

$$\beta = \phi^{c_8} \tag{25}$$

where $EGR$ is EGR fraction, $P_{SOI}$ and $T_{SOI}$ are the temperature and pressure at SOI, $X_d$ indicates the dilution fraction, $N$ denotes the engine speed, $\phi$ represents the equivalence ratio of the fuel, and $c_1$ to $c_9$ are the constants shown in Table 5. In the CA50 dynamic model shown in Eqns. (21) – (25), the states $x_1$ and $x_2$ change from cycle to cycle, and the parameters $\alpha$ and $\beta$ are calculated based the measurements of engine speed $N$ and equivalence ratio $\phi$.

## B. Adaptive Control System Structure

Based on the CA50 dynamic model shown in Eqns. (21) – (25), the control system is designed as shown in Fig. 9. An optimal CA50 is given as the reference CA50 and this reference along with engine speed, equivalence ratio and the actual CA50 serve as the inputs to the adaptive controller. Based on these inputs, the adaptive controller calculates the appropriate SOI that is then sent to the diesel engine.

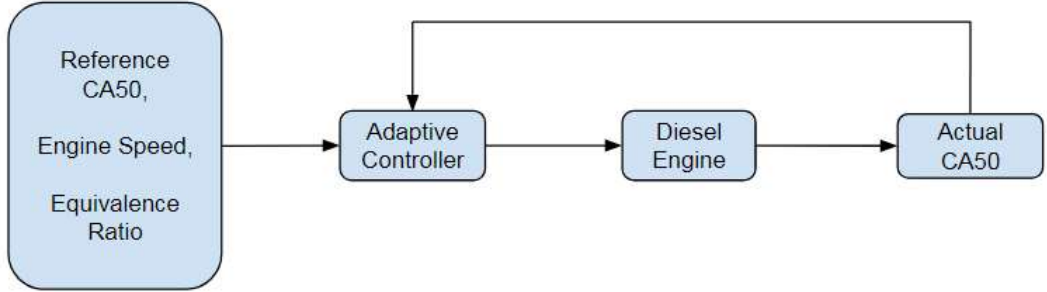

Figure 9. Structure of CA50 adaptive feedback control system

## C. Adaptive Controller Design

Using Eqn. (21), an appropriate input ($u$) to the control system can be found based on the parameters, states and the reference output $y_d$ (the desired CA50). The parameters $\alpha$ and $\beta$ can be computed using engine speed $N$ and equivalence ratio $\phi$. As such, the input of the control system can be given as:

$$u = y_d - \alpha x_1 - \beta x_2 \tag{26}$$

However, because the states $x_1$ and $x_2$ vary from cycle to cycle and they cannot be measured directly from sensors, an observer is needed to estimate their values. Therefore, Eqn. (26) can be rewritten as:

$$u = y_d - \alpha \bar{x}_1 - \beta \bar{x}_2 \tag{27}$$

in which $\bar{x}_1$ and $\bar{x}_2$ are the observed states.

In this observer, a gradient descent algorithm is applied to estimate the states. In order to design an observer that accurately estimates the actual output $y$, the RMSE is used. The RMSE can be written as:

$$E = \frac{1}{2}(y - \bar{y})^2 \tag{28}$$

where $E$ is the RMSE, $y$ denotes the actual output and $\bar{y}$ expresses the observed output. The observed output $\bar{y}$ can be captured by:

$$\bar{y} = u + \alpha \bar{x}_1 + \beta \bar{x}_2. \tag{29}$$

During steady state conditions, the parameters and states should be constant. Since the input $u$ is calculated by the parameters and observed states at the last cycle and the desired output $y_d$ has a constant value, the desired observed output $\bar{y}$ should be the same as the desired output $y_d$ in steady state conditions. Thus, Eqn. (29) can be written as:
$$E = \frac{1}{2}(y - y_d)^2. \qquad (30)$$
With the definition of the error function in Eqn. (30), the partial derivatives of the error can be derived as
$$\frac{\partial E}{\partial \bar{x}_1} = -\alpha(y - y_d) \qquad (31)$$
$$\frac{\partial E}{\partial \bar{x}_2} = -\beta(y - y_d) \qquad (32)$$
Using the gradient descent algorithm, the states can be updated from cycle to cycle by following equations:
$$\bar{x}_1(k+1) = \bar{x}_1(k) - \eta \frac{\partial E}{\partial \bar{x}_1} \qquad (33)$$
$$\bar{x}_2(k+1) = \bar{x}_2(k) - \eta \frac{\partial E}{\partial \bar{x}_2} \qquad (34)$$
where $\bar{x}_1(k+1)$, $\bar{x}_2(k+1)$, $\bar{x}_1(k)$ and $\bar{x}_2(k)$ are the observed states at the $k+1$ cycle and $k$ cycle, respectively; $\eta$ represents the learning rate of the algorithm; and $\frac{\partial E}{\partial \bar{x}_1}$ and $\frac{\partial E}{\partial \bar{x}_2}$ are the partial derivatives given in Eqns. (31) and (32).

Substituting (31) into (33) and substituting (32) into (34), the state update equations can be captured as
$$\bar{x}_1(k+1) = \bar{x}_1(k) + \eta\alpha(y - y_d) \qquad (35)$$
$$\bar{x}_2(k+1) = \bar{x}_2(k) + \eta\beta(y - y_d) \qquad (36)$$
In this work, the learning rate $\eta$ is given by
$$\eta = \frac{1}{\alpha^2 + \beta^2} \qquad (37)$$
and is chosen to achieve system stability, a short settling time and low overshoot. Substituting Eqn. (37) into Eqns. (35) and (36), the dynamic equations of the observer are
$$\bar{x}_1(k+1) = \bar{x}_1(k) + \frac{\alpha}{\alpha^2 + \beta^2}(y - y_d) \qquad (38)$$
$$\bar{x}_2(k+1) = \bar{x}_2(k) + \frac{\beta}{\alpha^2 + \beta^2}(y - y_d) \qquad (39)$$
The control system will leverage Eqns. (27), (38) and (39) to track the desired CA50.

## D. Proof of Adaptive Control System Stability

For a practical application, it is necessary to guarantee the stability of the control system. For this combustion phasing control method, a Lyapunov direct method is utilized to prove the stability of the system. Because the control system is designed to track the desired CA50, the Lyapanov function is chosen to be
$$V[x(k)] = (y_d - y)^2 \qquad (40)$$

The chosen Lyapunov function $V[x(k)]$ is 0 only if $y_d$ equals $y$ and $V[x(k)]$ will approach infinity if $y_d - y$ approaches infinity.

Substituting Eqns. (21) and (27) into Eqn. (40), the Lyapunov function can be expressed as

$$V[x(k)] = \left[\alpha(x_1(k) - \bar{x}_1(k)) + \beta(x_2(k) - \bar{x}_2(k))\right]^2. \tag{41}$$

To simplify the equation, the errors of state estimations are defined in Eqns. (42) and (43).

$$\tilde{x}_1(k) = x_1(k) - \bar{x}_1(k) \tag{42}$$

$$\tilde{x}_2(k) = x_2(k) - \bar{x}_2(k) \tag{43}$$

Substituting Eqns. (42) and (43) into Eqn. (41), the Lyapunov function can be written as

$$V[x(k)] = [\alpha\tilde{x}_1(k) + \beta\tilde{x}_2(k)]^2. \tag{44}$$

Similar to Eqn. (44), Lyapunov function at the $k+1$ cycle can be given as:

$$V[x(k+1)] = [\alpha\tilde{x}_1(k+1) + \beta\tilde{x}_2(k+1)]^2 \tag{45}$$

where

$$\tilde{x}_1(k+1) = x_1(k+1) - \bar{x}_1(k+1) \tag{46}$$

$$\tilde{x}_2(k+1) = x_2(k+1) - \bar{x}_2(k+1) \tag{47}$$

Substituting Eqns. (37) and (38) into Eqns. (46) and (47) respectively, the errors in the state estimates at the $k+1$ cycle can be captured by Eqns. (48) and (49).

$$\tilde{x}_1(k+1) = x_1(k+1) - \bar{x}_1(k) - \frac{\alpha}{\alpha^2 + \beta^2}(y - y_d) \tag{48}$$

$$\tilde{x}_2(k+1) = x_2(k+1) - \bar{x}_2(k) - \frac{\beta}{\alpha^2 + \beta^2}(y - y_d) \tag{49}$$

Taking the difference between Eqn. (21) and Eqn. (29) yields

$$y - \bar{y} = \alpha(x_1(k) - \bar{x}_1(k)) + \beta(x_2(k) - \bar{x}_2(k)) \tag{50}$$

Since the observed output $\bar{y}$ and reference output $y_d$ are the same, Eqn. (50) can be rewritten as

$$y - y_d = \alpha(x_1(k) - \bar{x}_1(k)) + \beta(x_2(k) - \bar{x}_2(k)). \tag{51}$$

Substituting Eqns. (42) and (43) into Eqn. (51), the equation becomes

$$y - y_d = \alpha\tilde{x}_1(k) + \beta\tilde{x}_2(k). \tag{52}$$

Then, substituting Eqn. (52) into Eqns. (48) and (49), the state estimations errors in Eqns. (48) and (49) can be given as:

$$\tilde{x}_1(k+1) = x_1(k+1) - \bar{x}_1(k) - \frac{\alpha}{\alpha^2 + \beta^2}[\alpha\tilde{x}_1(k) + \beta\tilde{x}_2(k)] \tag{53}$$

$$\tilde{x}_2(k+1) = x_2(k+1) - \bar{x}_2(k) - \frac{\beta}{\alpha^2 + \beta^2}[\alpha\tilde{x}_1(k) + \beta\tilde{x}_2(k)] \tag{54}$$

At steady state, the states in Eqn. (21) are constant such that

$$x_1(k+1) = x_1(k) \tag{55}$$

$$x_2(k+1) = x_2(k) \tag{56}$$

If Eqns. (55) and (56) are substituted into Eqns. (53) and (54) respectively, the state estimation errors can be given by

$$\tilde{x}_1(k+1) = x_1(k) - \bar{x}_1(k) - \frac{\alpha}{\alpha^2 + \beta^2}[\alpha\tilde{x}_1(k) + \beta\tilde{x}_2(k)] \tag{57}$$

$$\tilde{x}_2(k+1) = x_2(k) - \bar{x}_2(k) - \frac{\beta}{\alpha^2 + \beta^2}[\alpha\tilde{x}_1(k) + \beta\tilde{x}_2(k)] \tag{58}$$

Substituting Eqns. (42) and (43) into Eqns. (57) and (58), these equations are simplified as in Eqns. (59) and (60).

$$\tilde{x}_1(k+1) = \tilde{x}_1(k) - \frac{\alpha}{\alpha^2 + \beta^2}[\alpha\tilde{x}_1(k) + \beta\tilde{x}_2(k)] \tag{59}$$

$$\tilde{x}_2(k+1) = \tilde{x}_2(k) - \frac{\beta}{\alpha^2 + \beta^2}[\alpha\tilde{x}_1(k) + \beta\tilde{x}_2(k)] \tag{60}$$

The Lyapunov function can then be expressed as

$$V[x(k+1)] = \begin{bmatrix} \alpha\left[\tilde{x}_1(k) - \frac{\alpha}{\alpha^2 + \beta^2}[\alpha\tilde{x}_1(k) + \beta\tilde{x}_2(k)]\right] + \\ \beta\left[\tilde{x}_2(k) - \frac{\beta}{\alpha^2 + \beta^2}[\alpha\tilde{x}_1(k) + \beta\tilde{x}_2(k)]\right] \end{bmatrix}^2 \tag{61}$$

by substituting Eqns. (59) and (60) into Eqn. (45).

At steady state conditions, simplification of Eqn. (61) can show that

$$V[x(k+1)] = 0 \tag{62}$$

Therefore, the difference between the Lyapunov function at $k+1$ cycle and at $k$ cycle is given in Eqn. (66).

$$V[x(k+1)] - V[x(k)] = -(y_d - y)^2 \tag{63}$$

Based on Eqn. (63), the difference between the Lyapunov function at $k+1$ cycle and at $k$ cycle is negative if $y_d - y$ is not zero.

In summary, it is shown that:

$$\begin{cases} V[x(k)] = 0, \text{if } y_d - y = 0, \\ V[x(k)] > 0, \forall\, y_d - y \neq 0, \\ V[x(k)] \to \infty, \text{if } y_d - y \to \infty \text{ and} \\ V[x(k+1)] - V[x(k)] < 0, \forall\, y_d - y \neq 0. \end{cases} \tag{64}$$

According to the Lyapunov direct method [35], the output of control system is globally asymptotically stable. While this proves the stability of the adaptive controller, the impact of errors in the real engines also needs to be considered. If the CA50 feedback signal from the in-cylinder pressure trace exhibits large oscillations, oscillation in the parameter estimations may also occur. This phenomenon may lead to longer settling time, and more significant issues like erroneous signals could lead to an unstable system. As such, the accuracy and stability of this control technique is tied to that of the CA50 measurement.

# V. Feedforward Model-Based Controller Design

Because the adaptive feedback control strategy requires a feedback CA50 signal and this measurement is not available for most commercial diesel engines or reliable during transient conditions, a feedforward control strategy was considered that does not rely on in-cylinder sensors. For this purpose, a feedforward model-based control strategy is designed and a block diagram of this control strategy is given in Fig. 10.

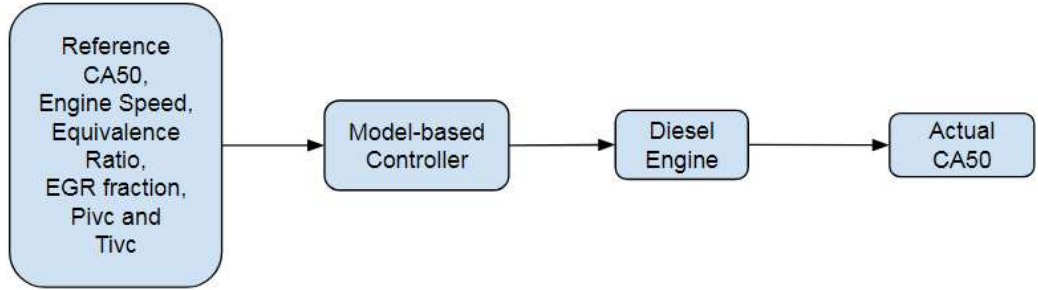

Figure 10. Structure of CA50 feedforward model-based control system

In this feedforward model-based control system, the reference CA50 is the optimal CA50. An appropriate SOI is given by the feedforward controller based on the reference CA50, engine speed, equivalence ratio, EGR fraction, pressure and temperature at IVC (intake valve closing). The fuel is injected into the diesel engine and produces an actual CA50 that is not measured.

As mentioned before, the CA50 prediction model is captured by Eqn. (20). Thus, the needed SOI can be calculated by

$$SOI = CA50_{ref} - (c_1 EGR + c_2)N\phi^{-c_3}\exp\left(\frac{c_4 P_{SOI}{}^{c_5}}{T_{SOI}}\right) \\ - c_9(1+X_d)^{c_7}\phi^{c_8} \qquad (65)$$

where $CA50_{ref}$ is the reference CA50. Although Eqn. (65) can be used to calculate the input SOI for the diesel engine combustion process, there are 3 parameters that cannot be measured directly namely the pressure and temperature at SOI and the dilution fraction. The pressure at SOI $P_{SOI}$ and the temperature at SOI $T_{SOI}$ can be found by Eqns. (16) and (17) using $P_{IVC}$ and $T_{IVC}$, and the dilution fraction $X_d$ can be determined based on Eqn. (6). Here the actual the actual $P_{IVC}$ and $T_{IVC}$ are used but a semi-empirical model for these variables could also be leveraged as in [20].

Therefore, substituting Eqns. (6), (16) and (17) into Eqn. (65), the SOI can be given as

$$SOI = CA50_{ref} - (c_1 EGR + c_2)N\phi^{-c_3}\exp\left[\frac{c_4\left[P_{IVC}\left(\frac{V_{IVC}}{V_{SOI}}\right)^{k_c}\right]^{c_5}}{T_{IVC}\left(\frac{V_{IVC}}{V_{SOI}}\right)^{k_c-1}}\right] \\ - c_9(1+EGR+X_r)^{c_7}\phi^{c_8} \qquad (66)$$

where $V_{SOI}$ is the volume of cylinder at SOI.

In Eqn. (66), 9 parameters are needed to predict SOI. Among these parameters, $EGR$ and $N$ can be taken directly from the ECU or from underlying estimators. Uncertainties in these variables can occur, particularly for EGR, and this will be explored in future work. Meanwhile, the equivalence ratio $\phi$ can be computed by Eqn. (3). The pressure and temperature of the intake manifold at IVC can estimated based on the intake manifold conditions. The value of $V_{\text{IVC}}$ is a function of cylinder geometry, which is known.

Besides these 7 parameters, $V_{\text{SOI}}$ and $X_r$ are needed but present a challenge. Since SOI is not known a priori, $V_{\text{SOI}}$ and SOI would have to be solved for iteratively. However, iteration leads to additional computation time for the ECU. From previous tests, it is known that the SOI should be close to 0 CAD to achieve the optimal CA50. Because the volume of cylinder does not change much when it is close to 0 CAD, the volume of cylinder at 0 CAD $V_0$ is applied to replace $V_{\text{SOI}}$. In addition, among the 516 simulations that are used to calibrate the CA50 model, the residual fraction $X_r$ only ranges from 0.02 to 0.06. Compared with the EGR fraction, $X_r$ is quite small. Therefore, the average residual fraction $\overline{X_r}$ ($\overline{X_r} = 0.0384$) is used in this model-based control. Engines with variable valve timing are able to trap additional residual gas and as such a more detailed model of residuals such as that in [20] may be needed.

Based on the analysis above, the SOI can be predicted by Eqn. (67).

$$SOI = CA50_{\text{ref}} - (c_1 EGR + c_2) N \phi^{-c_3} \exp\left[\frac{c_4 \left[P_{\text{IVC}} \left(\frac{V_{\text{IVC}}}{V_0}\right)^{k_c}\right]^{c_5}}{T_{\text{IVC}} \left(\frac{V_{\text{IVC}}}{V_0}\right)^{k_c-1}}\right] - c_9 (1 + EGR + \overline{X_r})^{c_7} \phi^{c_8} \quad (67)$$

With Eqn. (67), the combustion phasing control system with model-based feedforward control can decide on an appropriate SOI that should give the desired CA50.

With this feedforward model-based controller, the control error is highly dependent on the accuracy of the measurements. Measurement errors, signal noise or cyclic variations could all detrimentally affect the CA50 predictions. The error in CA50 prediction due to errors in $P_{\text{IVC}}$, $T_{\text{IVC}}$, EGR fraction, equivalence ratio and $X_r$ have been investigated. In [20], a semi-empirical model for $P_{\text{IVC}}$ and $T_{\text{IVC}}$ is proposed. The uncertainty with this model is ±0.036 bar for $P_{\text{IVC}}$, and the uncertainty of $T_{\text{IVC}}$ is ±3.8 K. For safety, ±0.05 bar error in $P_{\text{IVC}}$ and ±5 K error in $T_{\text{IVC}}$ are tested. Errors due to inaccurate EGR fraction, equivalence ratio and residual gas fraction estimates were also evaluated. The resulting loss in accuracy of the CA50 prediction was studied for all 516 simulation cases (which were used to validate the model). The resulting errors are listed in Table 6. As demonstrated in Table 6, the error in CA50 prediction is typically still small enough for the feedforward model-based controller design to work properly.

Table 6.  Error Response of CA50 Prediction

| Error Source | Error Value | Standard Deviation of CA50 Prediction Error (CAD) | Maximum of CA50 Prediction Error (CAD) |
|---|---|---|---|
| No Error | - | 0.33 | 0.87 |
| $P_{IVC}$ | +0.05 bar | 0.32 | 0.95 |
| $P_{IVC}$ | -0.05 bar | 0.33 | 0.93 |
| $T_{IVC}$ | +5K | 0.31 | 1.05 |
| $T_{IVC}$ | -5K | 0.35 | 1.09 |
| EGR | +5% | 0.33 | 0.94 |
| EGR | -5% | 0.32 | 0.97 |
| Equivalence Ratio | +0.05 | 0.33 | 0.88 |
| Equivalence Ratio | -0.05 | 0.33 | 0.90 |
| $X_r$ | +0.03 | 0.33 | 0.86 |
| $X_r$ | -0.03 | 0.33 | 0.91 |

# VI. Simulation and Analysis

In this section, the two controllers based on adaptive feedback control and feedforward model-based control are evaluated in simulations. Each control method is tested for five different transients. In the first 5 seconds of the transient, the system is kept at an initial steady state operating point, and the operating point is changed at the 5 second mark. The system then transitions to the second steady state operating point during the period from 5 to 10 seconds. The detailed settings for each case will be shown with the simulation result and analysis. In these simulations, the precision of SOI is considered to be 0.1 CAD.

## A. Case 1: Reference CA50 Change

In the first case, the controllers are evaluated during a change in reference CA50. The settings for this case are given in Table 7 and the results are shown in Fig. 11. In Fig. 11, the green line represents the actual CA50 achieved with adaptive control, and the blue curve indicates the CA50 with feedforward model-based control.

Table 7.  Settings of Case 1

| Quantity | First Operating Point | Second Operating Point |
|---|---|---|
| Engine Speed (RPM) | 1200 | 1200 |
| Average Temperature at Intake Manifold (K) | 300 | 300 |
| Average Pressure at Intake Manifold (bar) | 2 | 2 |
| Diesel Equivalence Ratio (-) | 0.7 | 0.7 |
| EGR fraction (%) | 25 | 25 |
| Reference CA50 (CAD) | 8 | 10 |

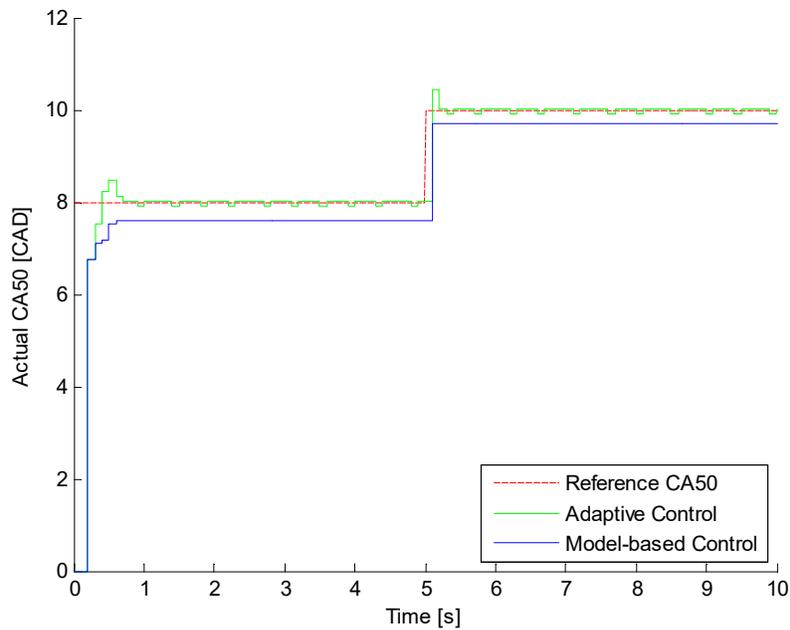

Figure 11.  Simulation result for Case 1

No fuel is injected into the cylinder in first 2 cycles, but after fuel injection begins, both control systems reach steady state in 5 cycles. The adaptive controller has a 0.46 CAD overshoot, and the steady state error ranges from -0.08 to 0.02 CAD. This oscillation of the error is mainly from the precision of SOI rather than the control algorithm. Compared with the adaptive controller, the feedforward model-based controller has a -0.40 CAD steady state error without overshoot. The feedforward model-based controller does not have such overshoot because it is an open-loop controller without the feedback iteration.

When the reference CA50 is changed, the actual CA50 is changed a cycle later. This is due to a delay in receiving the new reference CA50. At 5 seconds, the reference CA50 is still 8 CAD, and it changes to 10 CAD at 5.001 seconds. However, the controllers have already received the reference CA50 from the signal generator at 5 seconds. Therefore, the controllers decide on the SOIs based on the old reference CA50 of 8 CAD. After the reference change, the CA50 from the adaptive control gets back to the desired value in 2 cycles with a 0.46 CAD overshoot, and the steady state error is from -0.08 to 0.02 CAD. The feedforward model-based controller does not have overshoot, and the steady state error is -0.30 CAD.

## B. Case 2: Reference Engine Speed Change

The second test considers a change in engine speed. The settings in this case are listed in Table 8, and the simulation result is plotted in Fig. 12. Despite the engine speed changing at 5 seconds, the desired CA50 is still achieved. Again, there is no fuel input in the first 2 cycles and in the first 5 seconds, both control systems can reach their steady state values in 5 cycles. Because the first operating condition in Case 2 has the same settings with the first operating condition in Case 1, the overshoot and steady state error are identical. After 5 seconds, the engine speed moves from 1200 RPM to 1500 RPM. Both control methodologies do not cause any significant change during the transient, and they reach their steady state in 2 cycles. The steady state errors are -0.05 CAD and -0.47 CAD for the adaptive control system and feedforward model-based control system, respectively.

Table 8.     Settings of Case 2

| Quantity | First Operating Point | Second Operating Point |
|---|---|---|
| Engine Speed (RPM) | 1200 | 1500 |
| Average Temperature at Intake Manifold (K) | 300 | 300 |
| Average Pressure at Intake Manifold (bar) | 2 | 2 |
| Diesel Equivalence Ratio (-) | 0.7 | 0.7 |
| EGR fraction (%) | 25 | 25 |
| Reference CA50 (CAD) | 8 | 8 |

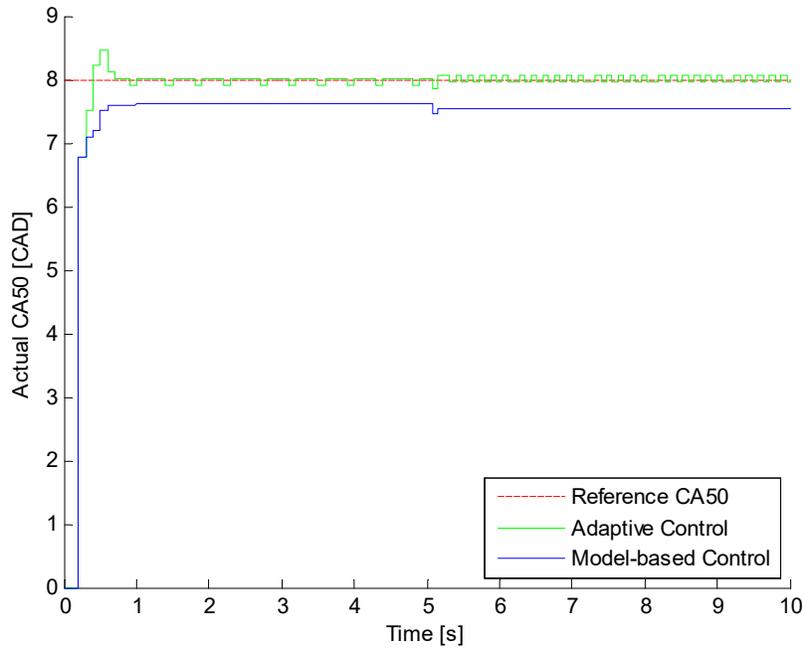

Figure 12. Simulation result for Case 2

## C. Case 3: Reference Intake Manifold Temperature Change

In Case 3, changes in the intake manifold temperatures are simulated as shown in Table 9.

Table 9.    Settings of Case 3

| Quantity | First Operating Point | Second Operating Point |
|---|---|---|
| Engine Speed (RPM) | 1200 | 1200 |
| Average Temperature at Intake Manifold (K) | 300 | 330 |
| Average Pressure at Intake Manifold (bar) | 2 | 2 |
| Diesel Equivalence Ratio (-) | 0.7 | 0.7 |
| EGR fraction (%) | 25 | 25 |
| Reference CA50 (CAD) | 8 | 8 |

Based on these parameters, the simulations are run in GT-ISE, and the result of the simulations are shown in Fig. 13.

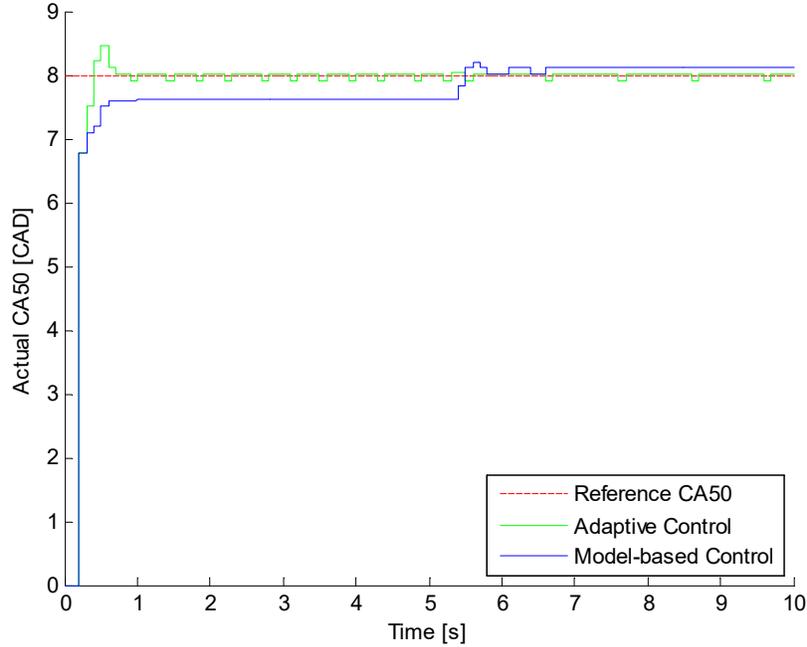

Figure 13.    Simulation result in Case 3

Like Case 1 and Case 2, no fuel injection in the first 2 cycles results in the zero CA50s in the first 2 cycles, but in the first 5 seconds, both control systems can reach steady quickly. The system based on adaptive control has an overshoot which is 0.46 CAD, and a steady state error from -0.08 CAD to 0.02 CAD. The system with feedforward model-based controller has a -0.40 CAD steady state error, which is larger than adaptive controller based system but still relatively low.

At 5 seconds, the average temperature of the intake manifold increases from 300 K to 330 K. The system with adaptive control does not have much change after the average temperature of the intake manifold changes and the steady state error is between -0.09 CAD and 0.01 CAD. On the contrary, the feedforward model-based control has more deviation during the transient. It has a maximum peak at 8.21 CAD and a 0.11 CAD steady state error. When the average intake manifold temperature has changed, it takes several cycles for the temperature at IVC $T_{IVC}$ to reach steady state. As seen in the nonlinear model calibration, the model is sensitive to the temperature changes. Therefore, when $T_{IVC}$ oscillates during the transient, the SOI given from the predictive model is also varied. This variation leads to the CA50 oscillation in the transient.

## D. Case 4: Reference Equivalence Ratio Change

In addition to considering changes in reference CA50s, engine speeds and average intake manifold temperature, changes in equivalence ratio are also evaluated in Case 4. The settings in Case 4 are listed in Table 10.

Table 10.    Settings of Case 4

| Quantity | First Operating Point | Second Operating Point |
|---|---|---|
| Engine Speed (RPM) | 1200 | 1200 |
| Average Temperature at Intake Manifold (K) | 300 | 300 |
| Average Pressure at Intake Manifold (bar) | 2 | 2 |
| Diesel Equivalence Ratio (-) | 0.5 | 0.9 |
| EGR fraction (%) | 25 | 25 |
| Reference CA50 (CAD) | 8 | 8 |

As seen in Fig. 14, in the first steady state condition, the steady state error oscillates from -0.05 CAD to 0.05 CAD with adaptive control, while the feedforward model-based control has a -0.37 CAD steady state error. At the 5 second mark, the diesel equivalence ratio jumps from 0.5 to 0.9. This change leads to more fuel being injected into the cylinder. The increasing fuel causes changes in the composition, temperature and pressure in the exhaust manifold. Since the diesel engine is integrated with EGR in this simulation, the pressure and temperature at the intake manifold also changes as shown in Fig. 15. These changes also take several cycles to get to their new steady state values. The steady state error with adaptive control is -0.07 – 0.04 CAD and the error from feedforward model-based control is -0.27 CAD.

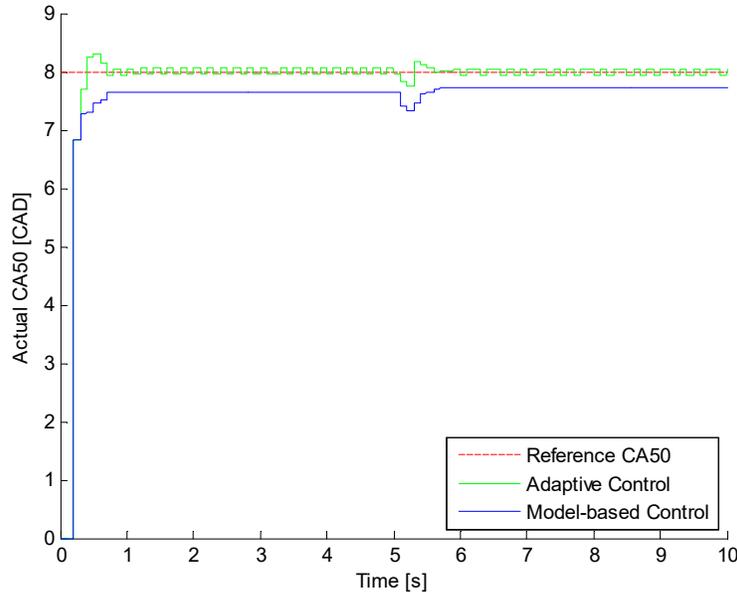

Figure 14. Simulation result in Case 4

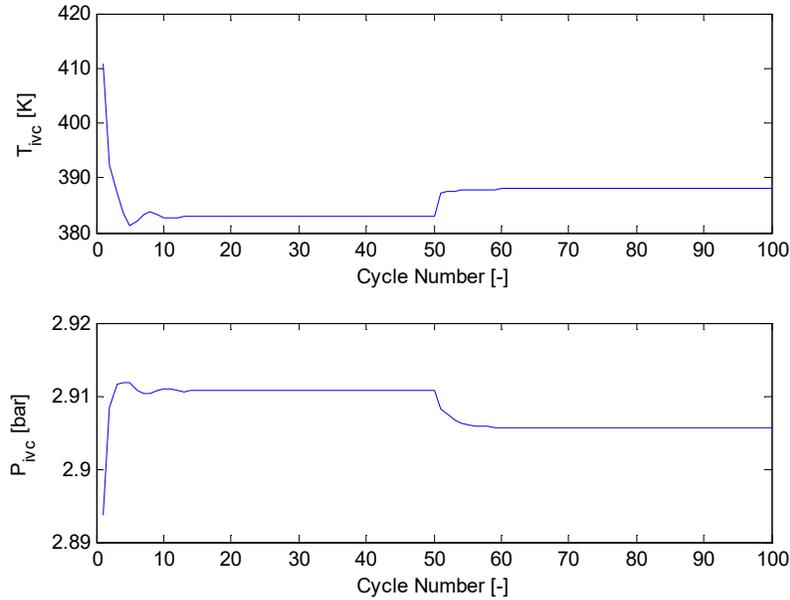
Figure 15. Pressure and temperature at IVC in Case 4

## E. Case 5: EGR Fraction Change

An EGR fraction change is tested in Case 5. Changes in EGR will result in changes to the temperature and pressure at IVC which in turn can disturb the CA50. The settings and simulation results are given in Table 11 and Fig. 16, respectively.

Table 11.     Settings of Case 5

| Quantity | First Operating Point | Second Operating Point |
|---|---|---|
| Engine Speed (RPM) | 1200 | 1200 |
| Average Temperature at Intake Manifold (K) | 300 | 300 |
| Average Pressure at Intake Manifold (bar) | 2 | 2 |
| Diesel Equivalence Ratio (-) | 0.7 | 0.7 |
| EGR fraction (%) | 0 | 50 |
| Reference CA50 (CAD) | 8 | 8 |

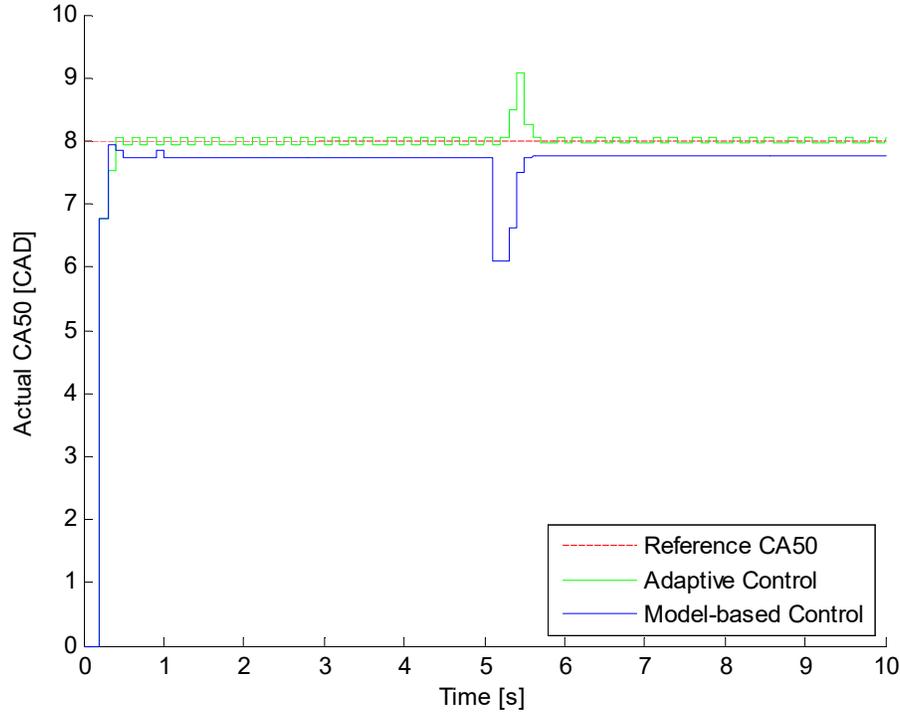

Figure 16. Simulation result in Case 5

Similar to the other 4 cases, zero fuel injection leads to no combustion in the first 2 cycles and both control systems reach the steady state after 5 cycles. The steady state error is -0.05 – 0.05 CAD for adaptive control, while the error for the feedforward model-based control is -0.26 CAD. The EGR fraction is changed from 0 to 0.5 after 5 seconds and both control approaches yield larger errors during the transient. The error is 1.07 CAD in adaptive control system, and it is -1.90 CAD for feedforward model-based control. As seen in Eqns. (22) and (23), the EGR fraction is a component in the states. When EGR fraction has such a jump from 0 to 0.5, the adaptive controller may need several cycles to track the change of EGR fraction and the states in Eqn. (24). Therefore, it has a large error during the transient. The properties at the intake manifold will also change due to the change of EGR fraction. As such, the pressure and temperature at IVC also change as shown in Fig. 17. Because the SOI given by the feedforward model-based control is calculated based on the parameters in the previous cycle, the feedforward model-based control cannot provide an accurate SOI when those parameters change rapidly. The steady state error is ranges from -0.04 CAD to 0.06 CAD with adaptive control and is -0.25 CAD with feedforward model-based control.

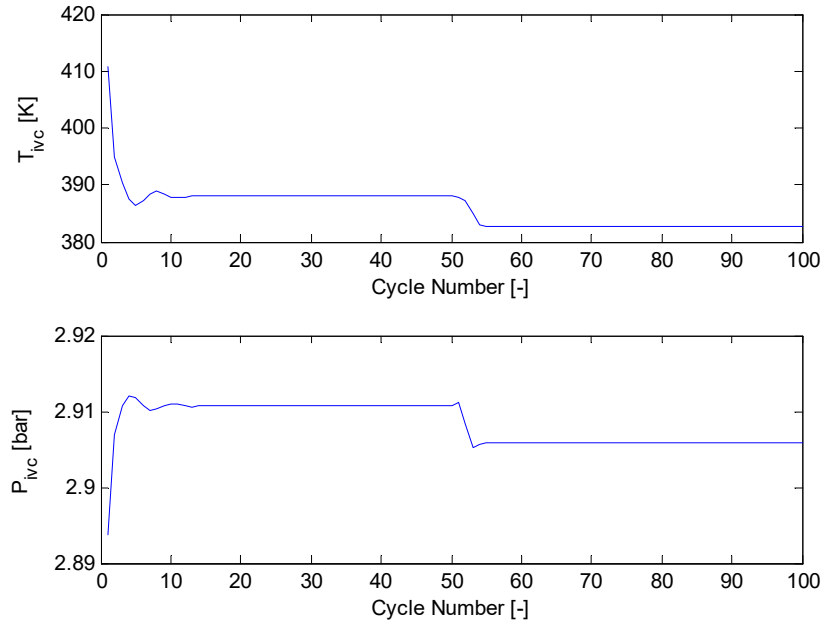

Figure 17. Pressure and temperature at IVC in Case 5

From the simulation results, it can be seen that both controllers can reach steady state quickly with a maximum steady state error less than ±0.1 CAD for adaptive control, and ±0.5 CAD for feedforward model-based control. However, both controllers have difficulty with tracking the reference CA50 when the parameters change significantly, especially in Case 5. However, a large EGR fraction change in 1 ms such as that used in Case 5 is not realistic and simply shows a worst-case scenario. The error during the transient would not be this large in a practical application. The adaptive controller is more accurate than the feedforward model-based controller, but CA50 measurements are not available for most of commercial diesel engines. Therefore, the system based on a feedforward model-based controller could be more widely used in stock diesel engines.

# VII. Conclusion

In this paper, MKIM, burn duration model and a Wiebe function are proposed to predict combustion phasing. These models are combined and simplified into a nonlinear model that is used to estimate CA50. Although MKIMs are widely used for the prediction of SOC and CA50 for different engines, it can be challenging to use them in the production engines because of the existence of the integral term. The simplification of the integral term makes it is much easier to predict and control CA50 in real engines. This non-linear model is calibrated and validated against simulations and experiments that include high EGR rates and high boost pressure. Interest in such conditions is growing as high dilution levels and high pressures are critical in many advanced combustion strategies. As such, this model could be useful for control in the conditions needed for high efficiency.

Based on this model, two control strategies are put forward. One leverages adaptive feedback control, and the other is a model-based feedforward control. The adaptive feedback control uses the actual CA50 measured from in-cylinder sensors and has a maximum steady state error less than ±0.1 CAD at different operating conditions. Since such in-cylinder sensors are not common for production engines, model-based control was also investigated. In this model-based control strategy, the average residual gas fraction is used instead of the dynamic residual gas fraction based on a complicated iteration. After this change, it becomes more feasible to leverage the model-based control strategy in the real engines. In the simulation tests, the combustion phasing control system with feedforward model-based controller has a maximum steady state error of less than ±0.5 CAD. Thus, both control strategies have a reasonable performance. Because the knock integral model and Wiebe function can be widely used in different engines, the control strategies developed in this paper could be applied to a variety of CI engines.

In future work, these strategies will be tested experimentally and the performance of the controllers during drive cycles will be also evaluated. Uncertainties in sensor measurements will also be considered in more detail in future work. Accurate measurements or estimates of EGR are also required for the proposed model and inaccuracies in such estimates will also be evaluated in the next stages of this work. This model can also be combined with a model of the gas exchange processes in order to provide a more accurate estimate of the temperature and pressure at IVC required by the combustion phasing model. Upcoming diesel technologies including low pressure EGR and variable valve timing could be directly integrated into this work since the EGR level and timing of valve closing are already variables in the model. However, if VVT is used to trap a significant amount of residuals, a more complex residual model may be needed to extend to these conditions.

# Acknowledgements

This material is based upon work supported by the National Science Foundation under Grant No. 1553823.

# Nomenclature

| Coefficient | Definition |
|---|---|
| $BD$ | Crank angle during burn duration |
| $CA50$ | Crank angle at 50% of fuel mass burnt |
| $CA50_{ref}$ | Reference CA50 |
| $EGR$ | Exhaust Gas Recirculation Fraction |
| $k_c$ | Polytropic constant |
| $m_{air}$ | Air mass entering the cylinder |
| $m_{egr}$ | EGR mass entering the cylinder |
| $m_{fuel}$ | Mass of injected fuel |
| $m_{residual}$ | Mass of residual gas in the cylinder |
| $N$ | Engine Speed |
| $P$ | In-cylinder dynamic pressure |
| $P_{IVC}$ | Pressure at intake valve close |
| $P_{SOI}$ | Pressure at start of injection |
| $SOC$ | Crank angle at start of combustion |
| $SOI$ | Crank angle at start of fuel injection |
| $T$ | In-cylinder dynamic temperature |
| $T_{IVC}$ | Temperature at intake valve close |
| $T_{SOI}$ | Temperature at start of injection |
| $V$ | Dynamic volume of cylinder |
| $V_0$ | Cylinder volume at 0 crank angle degree |
| $V_{IVC}$ | Cylinder volume at intake valve close |
| $V_{SOI}$ | Cylinder volume at start of injection |

| Coefficient | Definition |
|:---:|:---:|
| $x_b$ | Mass fraction of burnt fuel |
| $X_r$ | Mass fraction of residual gas |
| $X_d$ | Mass fraction of dilution |
| $\tau$ | Arrhenius function |
| $\phi$ | Diesel equivalence ratio |
| $\theta$ | Crank angle |

# Acronyms

| | |
|:---:|:---:|
| AFR | Air fuel ratio |
| aTDC | After top dead center |
| BD | Burn duration |
| CA50 | Crank angle at 50% fuel mass burnt |
| CAD | Crank angle degree |
| CFD | Computational fluid dynamics |
| ECU | Engine control unit |
| EGR | Exhaust gas recirculation |
| EVC | Exhaust valve close |
| EVO | Exhaust valve open |
| HCCI | Homogenous charge compression ignition |
| IVC | Intake valve close |
| IVO | Intake valve open |
| KIM | Knock integral model |
| LPPC | Location of peak premixed combustion |
| MKIM | Modified knock integral model |
| PI | Proportional–integral |
| PID | Proportional–integral–derivative |
| RBFNN | Radial basis function neural network |
| RMSE | Root mean squared error |
| RPM | Revolutions per minute |
| SOC | Start of combustion |
| SOI | Start of injection |
| st | Stoichiometric |
| VGT | Variable geometry turbocharger |